\documentclass[12pt,preprint]{aastex}

\newdimen\minuswidth    
\setbox0=\hbox{$-$}
\minuswidth=\wd0
\catcode`@=\active
\def@{\kern\minuswidth}
\newdimen\digitwidth    
\setbox0=\hbox{\rm0}
\digitwidth=\wd0
\catcode`!=\active
\def!{\kern\digitwidth}

\received{2004 October 28}
\begin{document}
\shorttitle{IR spectra of giant M stars}
\shortauthors{Rich \& Origlia}

\title{The First Detailed Abundances for M giants in Baade's Window from Infrared
Spectroscopy}
\altaffiltext{1}{
Data presented herein were obtained
at the W.M.Keck Observatory, which is operated as a scientific partnership
among the California Institute of Technology, the University of California,
and the National Aeronautics and Space Administration.
The Observatory was made possible by the generous financial support of the
W.M. Keck Foundation.}

\author{R. Michael Rich}
\affil{Physics and Astronomy Bldg,
430 Portola Plaza  Box 951547
Department of Physics and Astronomy, University of California
at Los Angeles, Los Angeles, CA 90095-1547\\
rmr@astro.ucla.edu}
\author{Livia Origlia}
\affil{INAF -- Osservatorio Astronomico di Bologna,
Via Ranzani 1, I--40127 Bologna, Italy,\\
livia.origlia@bo.astro.it}


\begin{abstract}
We report the first abundance analysis of 14 M giant stars in the Galactic
bulge, based on $R=25,000$ infrared spectroscopy  $(1.5-1.8 ~\mu \rm m)$ using NIRSPEC
at the Keck telescope.   Because some of the bulge M giants reach high luminosities
and have very late spectral type, it has been suggested that they are the progeny
of only the most metal rich bulge stars, or possibly members of a younger
bulge population.  We find the iron abundance and composition of the M giants
are similar to those of the K giants that have abundances determined from
optical high resolution spectroscopy: $\rm \langle [Fe/H] \rangle = -0.190 \pm 0.020$,
with a $1\sigma$ dispersion of $0.080\pm 0.015$.
Comparing our bulge M giants to a control sample of local disk M giants in the
solar vicinity, we find the bulge stars are enhanced in $\rm alpha$-elements at the level
of +0.3 dex relative the Solar composition stars, consistent with other studies of bulge globular
clusters and field stars.  This small sample shows no dependence of spectral type
on metallicity, nor is there any indication that the M giants are the evolved members of a subset
of the bulge population endowed with special characteristics such as relative youth
or high metallicity.
We also find
low $\rm ^{12}C/^{13}C\le10$, confirming the
presence of extra-mixing processes during the red giant phase of evolution.

\end{abstract}


\section{Introduction}
\label{intro}

The commissioning of the 4m telescope at Cerro Tololo (now the
Blanco telescope) included a new wide field grating/prism developed by
Blanco \citep{b01} at the prime focus,
permitting large scale surveys yielding slitless
spectra at classification resolution.
Used in combination with red sensitive photographic plates, the
grism was ideally suited to surveys of M and carbon stars;
the Magellanic Clouds and Galactic bulge were prime targets.
The discovery and spectral classification of hundreds of M giants
in the Baade's Window field \citep{bmb84} identified large numbers
of M giants with spectral types later than M4 and as cool as M9.  Such cool luminous
stars are not found in globular clusters, so this characteristic of the
bulge population was immediately understood at the time to
contradict the dominant view
of the bulge being a metal poor, population II system.  Using the
identifications of \citet{bmb84}, \citet{fw87} obtained infrared photometry from 
which they obtained effective temperatures and bolometric luminosities (reaching as bright as $M_{bol}<-4.5$) of the bulge
M giant population.

Even at high resolution, M giants are difficult to study using optical abundance
analysis due to their severe
molecular blanketing.   While \citet{MWR94} and subsequent studies perform
high resolution abundance analysis of
of the bulge K giants, the composition of
the Baade's Window M giants which includes stars as late as M9, 
continues to remain as an unsolved problem.   Because stars like these are a major
contributor to the integrated light of elliptical galaxies, the study of these stars has
broader implications.

During the 1980's, the indications of a wide abundance spread in the
bulge from the presence of both RR Lyrae and late M giant stars was 
confirmed from the study of the K giants \citep{Ri88}.
Photometry and low resolution spectroscopy of the K/M giant population 
indicated an average metallcity of [M/H]$\approx+0.3$\citep[see e.g.][]{Ri88,tfw90, TFW91}
and a larger spread ($\approx$1~dex) among K--giants compared to that of
the M giant population ($\approx$0.3~dex).  \citet{sharp90} found the bulge M giants to have 
greater TiO 8415 band line strength at constant $J-K$ relative to the Solar vicinity
stars; model atmosphere fits found a +0.5 dex [M/H] enhancement for the bulge giants.
Recently, the use of strong Ca, Na, and CO lines in the K band at medium spectral resolution
has been \citep{frog01} has been applied the bulge M giant population, finding roughly
Solar metallicity and no indication of a gradient in [M/H] within the inner 500 pc  \citep{ram00}.
For these cool stars especially, one would very much like to see abundances measured
from high resolution spectra as the gradient in composition remains unconstrained.
When the commissioning of NIRSPEC \citep{ml98} on Keck II made it possible to
obtain high dispersion infrared spectroscopy, we decided to 
obtain spectra of the Baade's Window M giants.
Having established abundances for the M giants in Baade's Window, our aim is to
extend our study to include the population of old late-type giants closer to the Galactic Center.

While all low mass stars ascend the first giant branch, only the
most metal rich stars might be expected to 
become M giants.  Are the M giants the
progeny of the most metal rich, or perhaps the most alpha-enhanced, bulge stars? 
Given their high luminosities, are the M giants be
a separate, perhaps younger, stellar population?  Detailed abundance
analysis plays a critical role in answering these questions.  Here we
report the first abundance analysis from high resolution infrared spectra,
 for M giants in the Galactic
bulge field of Baade's Window.

\section{Observations and abundance analysis}

\begin{figure}
\plotone{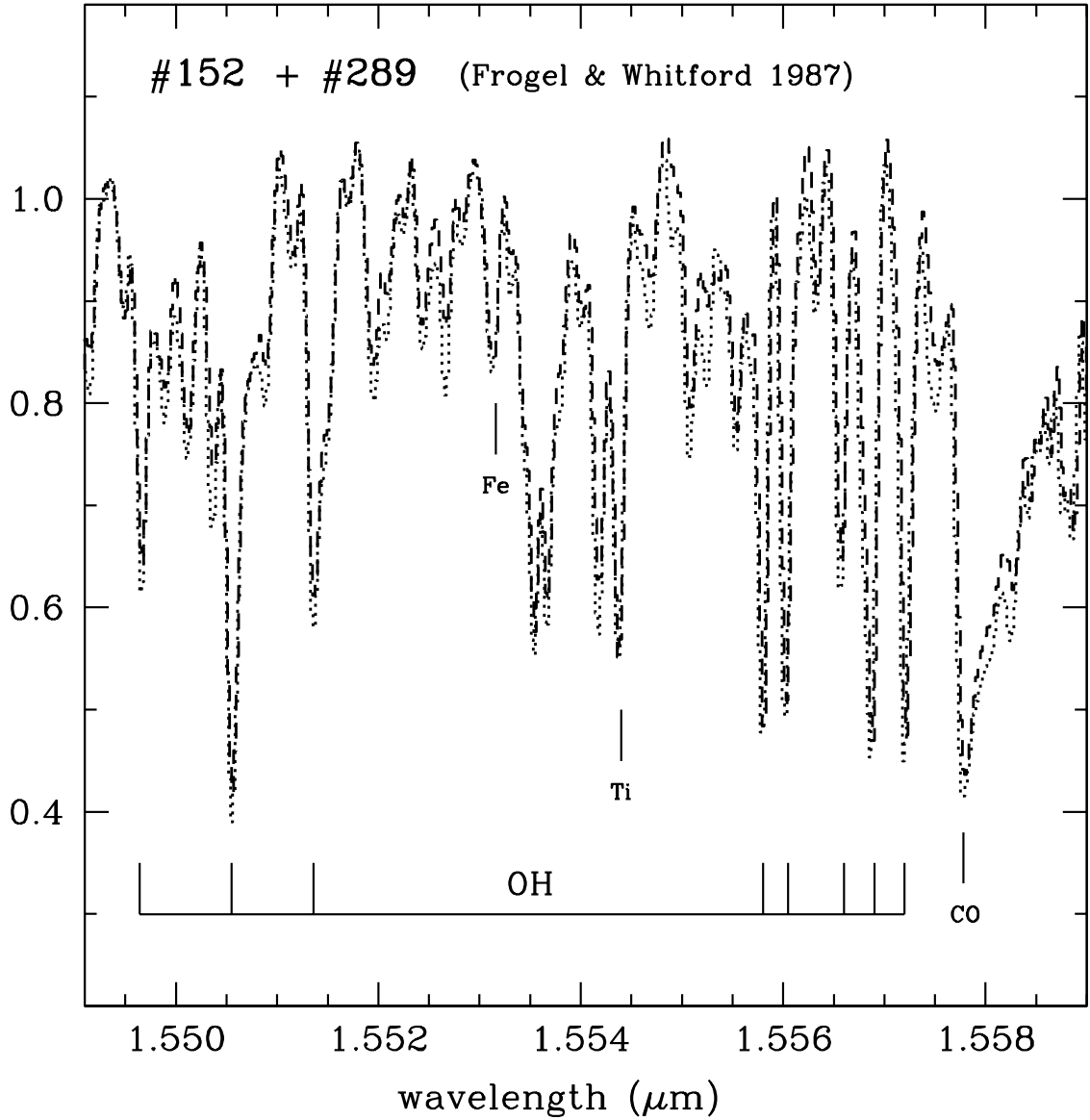}
\caption {We illustrate a section of the NIRSPEC H--band spectrum around 1.555~$\mu$m of
two M9 giants in the Baade's window.
A few major atomic lines and molecular bands of interest are
indicated.  BMB289 is the solid dashed line. \label{spec}}
\end{figure}
 
The Baade's Window field is a roughly 45$^\prime$ diameter circle centered on the globular cluster
NGC 6522 $(l,b)=(-3.93,1.02)$ corresponding to $\sim 550$pc
South of the Galactic Center, at 8kpc.   This field lies within the photometric structure of the
bulge/bar and represents well the bulge stellar population.   \citet{mould83} measured radial
velocities for 50 of the stars, finding a dispersion of $113\pm 11$ km/sec, a value consistent
with bulge membership and confirmed by \citet{sharp90} in their sample of 239
M giants.
The coordinates and finding charts for the
M giants are given in \citet{bmb84} and \citet{b86}; see Table 1).

The program stars were observed on 
 23 June 2001, 13-15 July 2002,
and 20-21 July 2003.  Typical exposure times for these bright (H=8-9) stars range from
120 to 240 sec.   The fainter K giants (I-195 and I-194) required
a total of 1800 sec of exposure time, as they are typically 3-4 mag fainter.
We nod the star by 5-10 arcsec to
take at least two separate exposures.  The M giants were frequently
observed in conditions of less than ideal seeing and transparency.
Four M giants in the Solar vicinity were observed on 27
April 2004.  We used NIRSPEC \citep{ml98} in the echelle mode.
A slit width of $0\farcs43$ and a length of 8\arcsec\ or 12\arcsec\ 
(depending on the targets) giving an overall spectral resolution R=25,000,
and the standard NIRSPEC-5 setting, which
covers most of the 1.5-1.8 $\mu$m H-band
have been selected. 

The raw stellar spectra have been reduced using the
REDSPEC IDL-based package written
at the UCLA IR Laboratory.
Each order has been
sky subtracted by using nodding pairs and flat-field corrected.
Wavelength calibration has been performed using arc lamps and a 2-nd order
polynomial solution, while telluric features have been removed by using
a O-star featureless spectrum.
The signal to noise ratio of the final spectra is $\ge$50. 
Fig.~\ref{spec} illustrates the spectrum in  the 1.555$\mu$m region for two M9 giants in our
sample.

A grid of suitable synthetic spectra
of giant stars has been computed 
by varying the photospheric parameters and the
element abundances, using an updated
version of the code described in \citet{OMO93}.
By combining full spectral synthesis analysis with equivalent widths
measurements of selected lines,
we derive  abundances for Fe and O, and with somewhat greater
errors, C, Mg, Si, and Ti.  The lines and analysis method are
described in \citet{orc02,ori04}.
Here we summarize the major issues.

The code uses the LTE approximation.
In the H band, most of the OH and CO molecular lines are not saturated and
can be safely treated under the LTE approximation, being roto-vibrational
transitions in the ground electronic state, providing accurate C and O abundances
\citep{mr79,lam84,sm00}.
Detailed computations of possible NLTE effects for atomic
lines in the H band have been performed only for AlI lines 
in the Sun (see Baumueller \& Gehren (1996), finding indeed negligible 
corrections.  However,
most of the near IR atomic lines are of 
high excitation potential, indicating that they form 
deep in the atmosphere, where the LTE approximation should hold even in 
giants of low gravity.
Moreover, one of the major mechanisms which can cause a 
deviation from LTE, namely over-ionization by UV radiation, is 
less efficient in cool giants, while photon suction 
can have some relevance.
According to NLTE computations on Fe and Mg lines 
\citep[see e.g.][]{gra99,zha00} 
deviations from LTE (at a level of $\ge$0.1~dex) 
are mainly observed in stars which are significantly hotter
and more metal poor than those in our program. 

\begin{figure}
\plotone{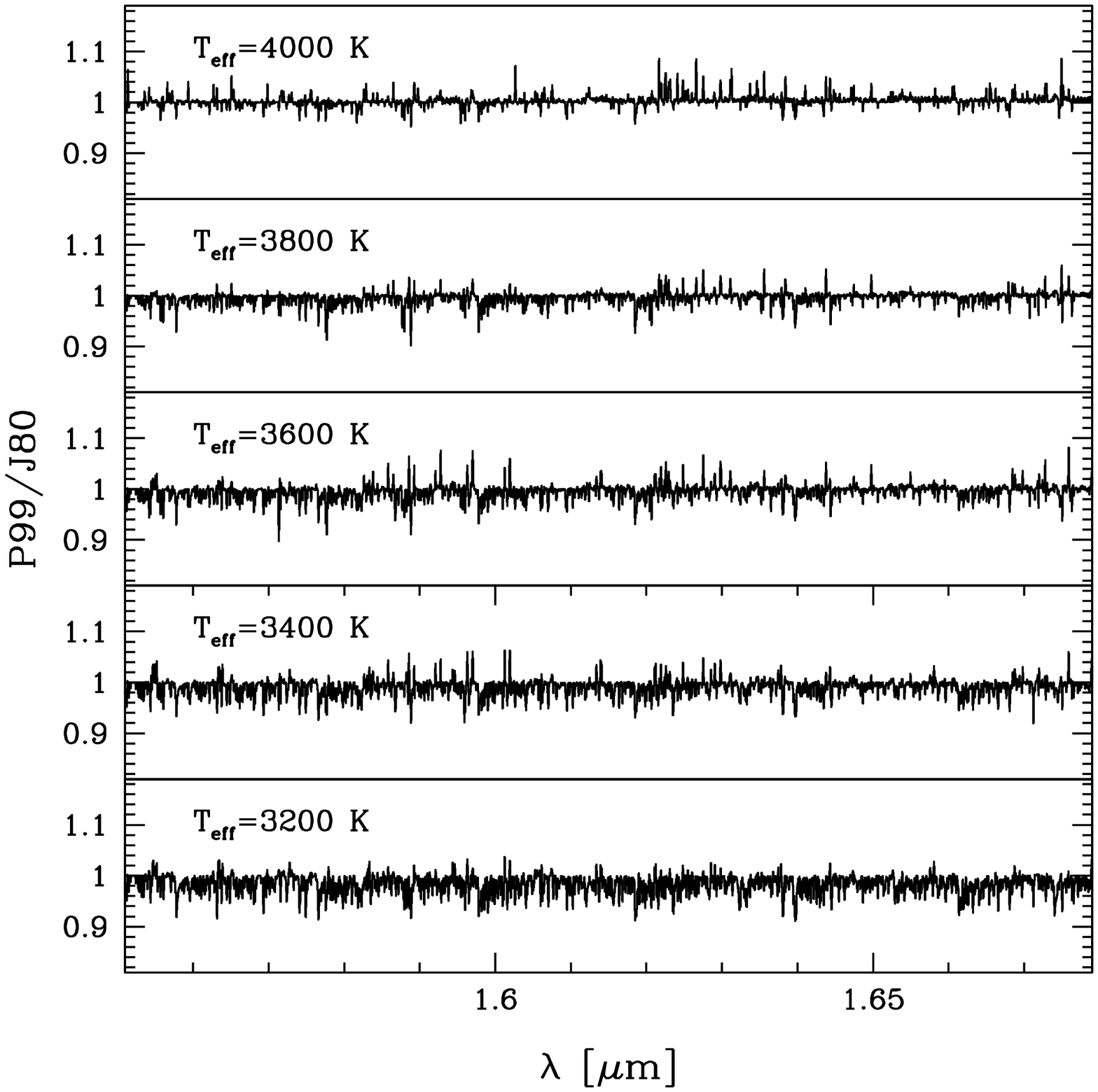}
\caption {Ratio between H band spectra at solar metallicity 
obtained using NextGen model atmospheres by \citet{hau99} and \citet{jbk80} ones 
for different temperatures which span fully the range of temperatures of our
sample.  For the 3200K case, the ratio of the Nextgen/J80 synthetic spectrum is
0.9832, with $\sigma=0.0177$ and a full range of 0.9116 to 1.0371.  We used the 3600K
nextgen atmosphere to analyse BMB93 and find that [Fe/H] and abundance ratios of
O, Si, Mg, Ca, Ti, and C/Fe agree to within 0.01 dex. }
\label{diffmod}
\end{figure}

The code is based
on the molecular blanketed model atmospheres of
\citet{jbk80} in the 3000-4000~K temperature range
and the ATLAS9 models for temperatures above 4000~K.
Since in the near IR
the major source of continuum opacity is H$^-$
with its minimum near 1.6 $\mu$m,
the dependence of the results on the choice of reasonable model
atmospheres should not be critical.
However, ss a check, we also computed synthetic spectra using the more updated 
NextGen model atmospheres by \citet{hau99} and we compare them with those obtained 
using \citet{jbk80} models. 
As an expample, Fig.~\ref{diffmod} shows the results at solar metallicity 
in the range of temperatures 3200-4000 K.
The average ratio between NextGen and \citet{jbk80} spectra is between 0.98 
(at T$_{eff}$=3200 K) 
and 1.0 (at T$_{eff}\ge$3600 K) over the entire H band, the most discrepant values  
being always within 0.9 and 1.1.
These discrepancies are most likely due to differences in the temperature structure of 
the model atmospheres which however have a minor impact on the overall inferred 
abundances.  To test this, we analyzed the 3600K M giant BMB93 using both atmospheres,
finding numerical results for [Fe/H] and O, Si, M, Ca, Ti, and C/Fe in agreement to within 0.02 
dex.  Other systematic errors are of course of much greater concern.

Three main compilations of
atomic oscillator strengths are used:
the Kurucz database
(c.f. {\it http://cfa-www.harward.edu/amdata/ampdata/kurucz23/sekur.html}),
\citet{bg73} and \citet{mb99}.
On average, log-gf values from \citet{mb99}
are systematically lower than Kurucz ones but
in most cases the difference does not exceed 0.2~dex
and the overall scatter in the derived abundances is $<0.1$~dex
\citep[see][for a more quantitative comparison]{orc02}.  The log-gf determinations
for the IR are not as good as for the optical and for this (and many other reasons)
we obtain spectra of local disk comparison M giants.

Applying this procedure to the IR spectrum of Arcturus \citep{hin95}, 
rebinned at the NIRSPEC resolution, we find abundances of 
[Fe/H]=$-0.6$, [O/H]=$-0.25$, [Mg,Si/H]=$-0.1$, [Ca/H]=$-0.45$, [Ti/H]=$-0.4$, 
[C/H]=$-0.8$, fully consistent (on average, within $\pm$0.1~dex) with the values published 
in the literature \citep{gw98,sm00,zoc04}.
Reference solar abundances are from \citet{gv98}.

In the first iteration, we estimate stellar temperature
from the $\rm (J-K)_0$ colors (see Table~\ref{tab1})
and the color-temperature transformation
of \citet{MFFO98} specifically calibrated on globular cluster giants.
Gravity has been estimated from theoretical evolutionary tracks,
according to the location of the stars on the Red Giant Branch (RGB)
\citep[see][and references therein for a more detailed
discussion]{ori97}.
For microturbulence velocity an average value
$\xi$=2.0 km/s has been adopted
\citep[see also][]{ori97}.
More stringent constraints on the stellar parameters are obtained by the
simultaneous spectral fitting of the several CO and OH molecular bands,
which are very sensitive to temperature, gravity and microturbulence variations 
(see Figs. 6,7 of \citet{orc02}). 

CO and OH in particular, are extremely sensitive to $T_{eff}$ in the range 
between 4500 and 3500 K.
Indeed, temperature sets the fraction of molecular {\it versus} atomic
carbon and oxygen.
At temperatures $\ge$4500 K molecules barely survive, most of the
carbon and oxygen are in atomic form and the CO and OH spectral features
become very weak.
On the contrary, at temperatures $\le$3500 K most of the carbon and oxygen
are in molecular form, drastically reducing the dependence of the CO and OH
band strengths and equivalent widths on the temperature itself
\citep{ori97}. 

\begin{deluxetable}{llclcrlllllllc}
\tabletypesize{\scriptsize}
\tablecaption{Stellar parameters and abundances for our sample of 
giant stars in Baade's window and the Solar neighborhood. \label{tab1}}
\tablewidth{0pt}
\tablehead{
\colhead{Star} & 
\colhead{SpT}&         
\colhead{(J-K)$_0$}&         
\colhead{$\rm T_{eff}$}&
\colhead{log~g}&
\colhead{$\rm v_r^c$}&
\colhead{$\rm [Fe/H]$}& 
\colhead{$\rm [O/Fe]$}& 
\colhead{$\rm [Si/Fe]$}& 
\colhead{$\rm [Mg/Fe]$}& 
\colhead{$\rm [Ca/Fe]$}& 
\colhead{$\rm [Ti/Fe]$}& 
\colhead{$\rm [C/Fe]$}& 
\colhead{$\rm ^{12}C/^{13}C$} 
}
\startdata
\multicolumn{14}{l}{\small Baade's window$^a$}\\
\multicolumn{14}{l}{}\\
BMB152  &M9& 1.32 &3200& 0.5 &  -22  &-0.24  & +0.41  & +0.24  & +0.44  & +0.35  & +0.24  &--0.26 &6.3\\
  &&& && &$\pm$0.07  &$\pm$0.11  &$\pm$0.16  &$\pm$0.20  &$\pm$0.12  &$\pm$0.14  &$\pm$0.12 &$\pm$1.5\\
BMB289  &M9& 1.37 &3200& 0.5 &  -83  &-0.15  & +0.26  & +0.15  & +0.15  & +0.25  & +0.15  &--0.35 &5.0\\
  &&& && &$\pm$0.06  &$\pm$0.10  &$\pm$0.15  &$\pm$0.13  &$\pm$0.11  &$\pm$0.14  &$\pm$0.12 &$\pm$1.2\\
BMB55   &M8& 1.30 &3200& 0.5 &   10  &-0.17  & +0.36  & +0.27  & +0.37  & +0.27  & +0.22  &--0.33 &5.6\\
  &&& && &$\pm$0.07  &$\pm$0.11  &$\pm$0.16  &$\pm$0.20  &$\pm$0.12  &$\pm$0.14  &$\pm$0.12 &$\pm$1.3\\
BMB165  &M7& 1.20 &3400& 0.5 &  133  &-0.29  & +0.35  & +0.29  & +0.36  & +0.39  & +0.40  &--0.31 &4.5\\
  &&& && &$\pm$0.08  &$\pm$0.12  &$\pm$0.16  &$\pm$0.15  &$\pm$0.12  &$\pm$0.14  &$\pm$0.13 &$\pm$1.1\\
BMB28   &M7& 1.24 &3400& 0.5 &  -79  &-0.22  & +0.36  & +0.32  & +0.27  & +0.32  & +0.21  &--0.28 &5.6\\
  &&& && &$\pm$0.06  &$\pm$0.15  &$\pm$0.20  &$\pm$0.15  &$\pm$0.11  &$\pm$0.15  &$\pm$0.12 &$\pm$1.3\\
BMB93   &M6& 1.05 &3600& 0.5 &  -66  &-0.15  & +0.31  & +0.25  & +0.30  & +0.25  & +0.21  &--0.35 &6.3\\
  &&& && &$\pm$0.08  &$\pm$0.10  &$\pm$0.16  &$\pm$0.14  &$\pm$0.12  &$\pm$0.14  &$\pm$0.13 &$\pm$1.5\\
BMB124  &M6& 1.06 &3600& 0.5 &  120  &-0.15  & +0.25  & +0.35  & +0.26  & +0.25  & +0.25  &--0.35 &5.6\\
  &&& && &$\pm$0.06  &$\pm$0.10  &$\pm$0.19  &$\pm$0.14  &$\pm$0.11  &$\pm$0.13  &$\pm$0.12 &$\pm$1.3\\
BMB133  &M6& 1.06 &3600& 0.5 &   89  &-0.23  & +0.30  & +0.33  & +0.38  & +0.33  & +0.33  &--0.27 &5.0\\
  &&& && &$\pm$0.07  &$\pm$0.11  &$\pm$0.16  &$\pm$0.14  &$\pm$0.12  &$\pm$0.14  &$\pm$0.12 &$\pm$1.2\\
B47     &M5& 1.00 &3800& 0.5 &  -78  &-0.20  & +0.27  & +0.27  & +0.30  & +0.30  & +0.35  &--0.25 &5.6\\
  &&& && &$\pm$0.07  &$\pm$0.08  &$\pm$0.16  &$\pm$0.20  &$\pm$0.11  &$\pm$0.12  &$\pm$0.12 &$\pm$1.3\\
B64     &M3& 0.94 &4000& 1.0 &   90  &-0.33  & +0.43  & +0.43  & +0.40  & +0.33  & +0.43  &--0.17 &6.3\\
  &&& && &$\pm$0.08  &$\pm$0.12  &$\pm$0.19  &$\pm$0.14  &$\pm$0.13  &$\pm$0.14  &$\pm$0.13 &$\pm$1.5\\
B66     &M3& 0.84 &4000& 1.0 &  -47  &-0.09  & +0.35  & +0.29  & +0.36  & +0.29  & +0.29  &--0.31 &5.0\\
  &&& && &$\pm$0.07  &$\pm$0.08  &$\pm$0.19  &$\pm$0.14  &$\pm$0.12  &$\pm$0.13  &$\pm$0.12 &$\pm$1.2\\
B78     &M2& 0.84 &4000& 1.0 &  -69  &-0.03  & +0.41  & +0.42  & +0.30  & +0.38  & +0.36  &--0.17 &7.0\\
  &&& && &$\pm$0.08  &$\pm$0.10  &$\pm$0.18  &$\pm$0.15  &$\pm$0.13  &$\pm$0.14  &$\pm$0.13 &$\pm$1.6\\
I195    &--& 0.98 &3800& 0.5 &  -13  &-0.25  & +0.22  & +0.25  & +0.16  & +0.35  & +0.33  &--0.35 &5.0\\
  &&& && &$\pm$0.07  &$\pm$0.09  &$\pm$0.22  &$\pm$0.18  &$\pm$0.11  &$\pm$0.13  &$\pm$0.12 &$\pm$1.2\\
I194    &--& 0.68 &4250& 1.5 & -180  &-0.13  & +0.22  & +0.33  & +0.23  & +0.33  & +0.13  &--0.37 &5.0\\
  &&& && &$\pm$0.09  &$\pm$0.10  &$\pm$0.18  &$\pm$0.19  &$\pm$0.16  &$\pm$0.14  &$\pm$0.13 &$\pm$1.2\\
\multicolumn{14}{l}{\small Solar neighborhood$^b$}\\
\multicolumn{14}{l}{}\\
BD--012971 &M5& 1.09 &3600& 0.5 &-120 &-0.24  &+0.04  & +0.04  & +0.05  & +0.04  & -0.02  &--0.46 &7.9\\
  &&& && &$\pm$0.07  &$\pm$0.09  &$\pm$0.19  &$\pm$0.15  &$\pm$0.10  &$\pm$0.13  &$\pm$0.12 &$\pm$1.8\\
BD+422310 &M2& 1.08 &3600& 0.5 &-50 &-0.11  & +0.04  & +0.01  & +0.06  & +0.06  & +0.01  &--0.59 &9.5\\
  &&& && &$\pm$0.07  &$\pm$0.10  &$\pm$0.19  &$\pm$0.14  &$\pm$0.11  &$\pm$0.14  &$\pm$0.14 &$\pm$2.0\\
BD+432283 &M4& 1.13 &3600& 0.5 &-68 &-0.14  & +0.04  & +0.04  & +0.04  & +0.07  & +0.04  &--0.56 &10\\
  &&& && &$\pm$0.09  &$\pm$0.11  &$\pm$0.19  &$\pm$0.15  &$\pm$0.13  &$\pm$0.15  &$\pm$0.12 &$\pm$2.3\\
BD+452076 &M3& 1.06 &3800& 0.5 &-60 &-0.30  & +0.12  & +0.10  & +0.09  & +0.10  & +0.10  &--0.39 &7.9\\
  &&& && &$\pm$0.07  &$\pm$0.10  &$\pm$0.19  &$\pm$0.15  &$\pm$0.11  &$\pm$0.12  &$\pm$0.12 &$\pm$1.8\\
\enddata                       
\tablenotetext{a}{Name and spectral type: BMB prefix from \citet{bmb84}; B prefix from \citet{b86}
and I prefix (K giants) from \citet{arp65};
photometry from \citet{FWR84,fw87,TFW91}.
The (J-K) color of I195 is from 2MASS and has been corrected for reddening adopting
E(J-K)=0.26 \citep{TFW91}.}
\tablenotetext{b}{The (J--K) colors are from 2MASS and have been corrected 
for reddening using \citet{sch98} extinction maps.} 
\tablenotetext{c}{Heliocentric radial velocity in $\rm km~s^{-1}$.}
\end{deluxetable}

For our analysis we compute a grid of synthetic spectra with
abundances and abundance patterns varying over a large range and
the photospheric parameters around the photometric estimates.
The final values of our best-fit models together with random errors
are listed in Table~\ref{tab1}.

\begin{figure}
\plotone{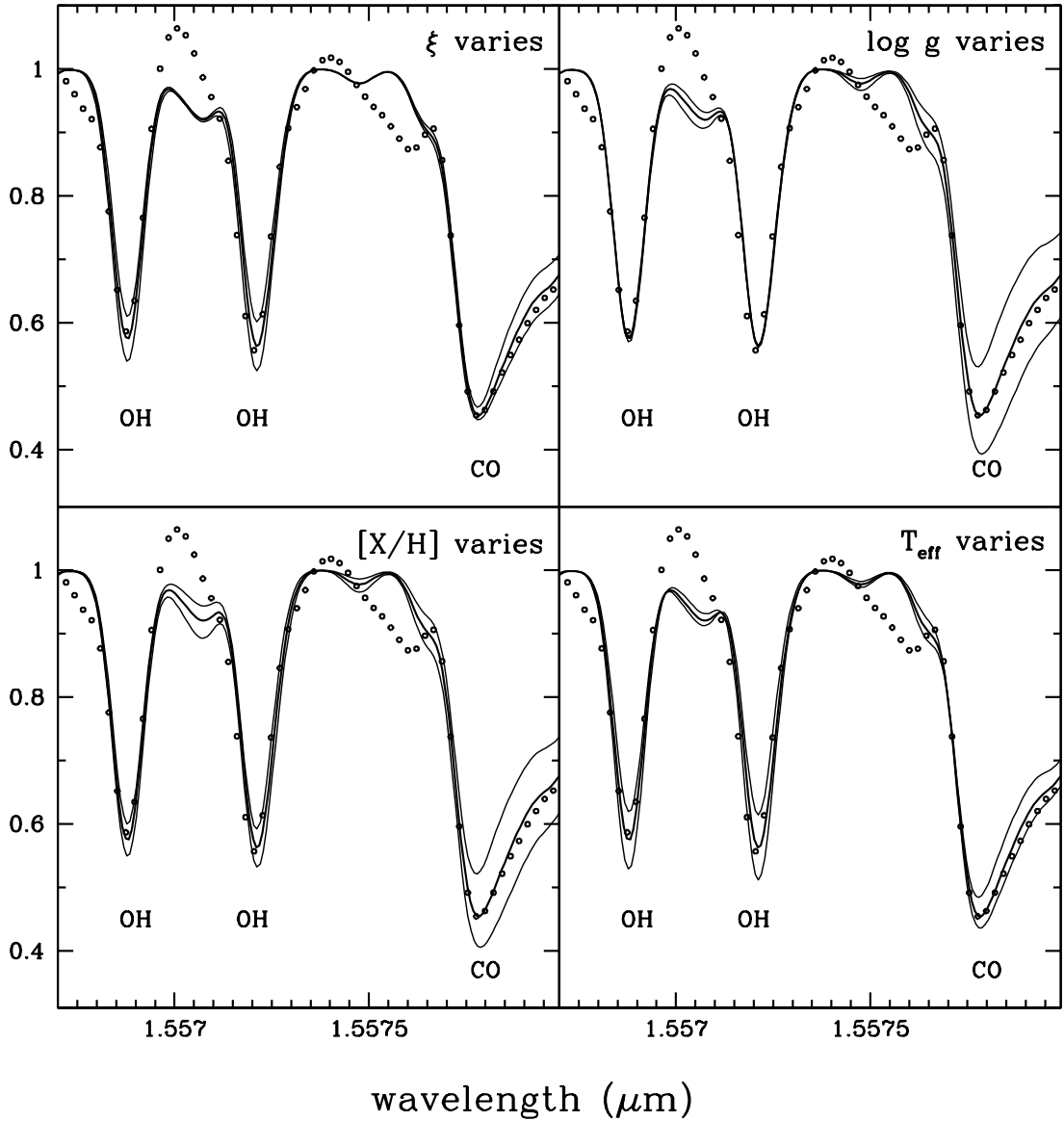}
\caption {Section of the H band spectruum of BMB93 (dots)
and our best fit (solid line), using
T$_{\rm eff}$=3600~K, log~g=0.5, $\xi$=2~km~s$^{-1}$, [Fe/H]=--0.2,
[O/Fe]=+0.3, [C/Fe]=--0.3 as reference stellar parameters 
(see also Table~\ref{tab1}).
For comparison we also plot synthetic spectra with 
different abundances and stellar parameters with respect 
to the best-fit solution. 
Bottom-left: $\rm \Delta [X/H]=\pm$0.2~dex; 
bottom-right: $\rm \Delta T_{eff}=\mp$200~K; 
top-left: $\rm \Delta \xi=\pm$0.5~km~s$^{-1}$;
top-right: $\rm \Delta log~g=\mp$0.5~dex.
}
\label{param}
\end{figure}

To illustrate,  Fig.~\ref{param} shows a region of the observed H band spectrum of BMB93
with (superimposed) our best-fit model and two other models with  
$\rm \Delta [X/H]=\pm$0.2~dex,
$\rm \Delta T_{eff}=\pm$200~K, 
$\rm \Delta \xi=\mp$0.5~km~s$^{-1}$, 
and $\rm \Delta log~g=\pm$0.5~dex, 
with respect to the best-fit parameters.
It is clearly seen that our models with $\pm$0.2~dex abundance or $\pm$200~K temperature 
variations give remarkably  different molecular line profiles.
Microturbulence variation of $\pm$0.5~km/s mainly affects the OH lines, while 
gravity mainly affects the CO lines.

As a further check of the statistical significance of our best-fit solution,
we also compute synthetic spectra with
$\rm \Delta T_{eff}=\pm$200~K, $\rm \Delta log~g=\pm$0.5~dex and
$\rm \Delta \xi=\mp$0.5~km~s$^{-1}$, and with corresponding simultaneous variations
of the C and O abundances (on average, $\pm$0.2~dex) to reproduce the depth of the 
molecular features.
As a figure of merit of the statistical test we adopt
the difference between the model and the observed spectrum (hereafter $\delta$).
In order to quantify systematic discrepancies, this parameter is
more powerful than the classical $\chi ^2$ test, which is instead
equally sensitive to {\em random} and {\em systematic} errors
\citep[see also][]{ori03,ori04}.
Our best fit solutions always show $>$90\% probability
to be representative of the observed spectra, while
spectral fitting solutions with abundance variations of $\pm$0.2~dex,
due to possible systematic uncertainties of $\pm$200~K in temperature,
$\pm$0.5~dex in gravity or $\mp$0.5 km/s in
microturbulence have always $\le$30\% probability to be
statistical significant.
Hence, as a conservative estimate of the systematic error in the derived 
best-fit abundances,
due to the residual uncertainty in the adopted stellar parameters, one can
assume a value of $\le \pm 0.1$~dex.
However, it must be noted that since the stellar features under consideration 
show a similar trend with variations in the stellar parameters, although with different
sensitivities, {\it relative } abundances are less
dependent on the adopted stellar parameters (i.e. on the systematic errors)
and their values are well constrained down to $\approx \pm$0.1~dex
(see also Table~\ref{tab1}).

\section{Results and Discussion}

\begin{figure}
\plotone{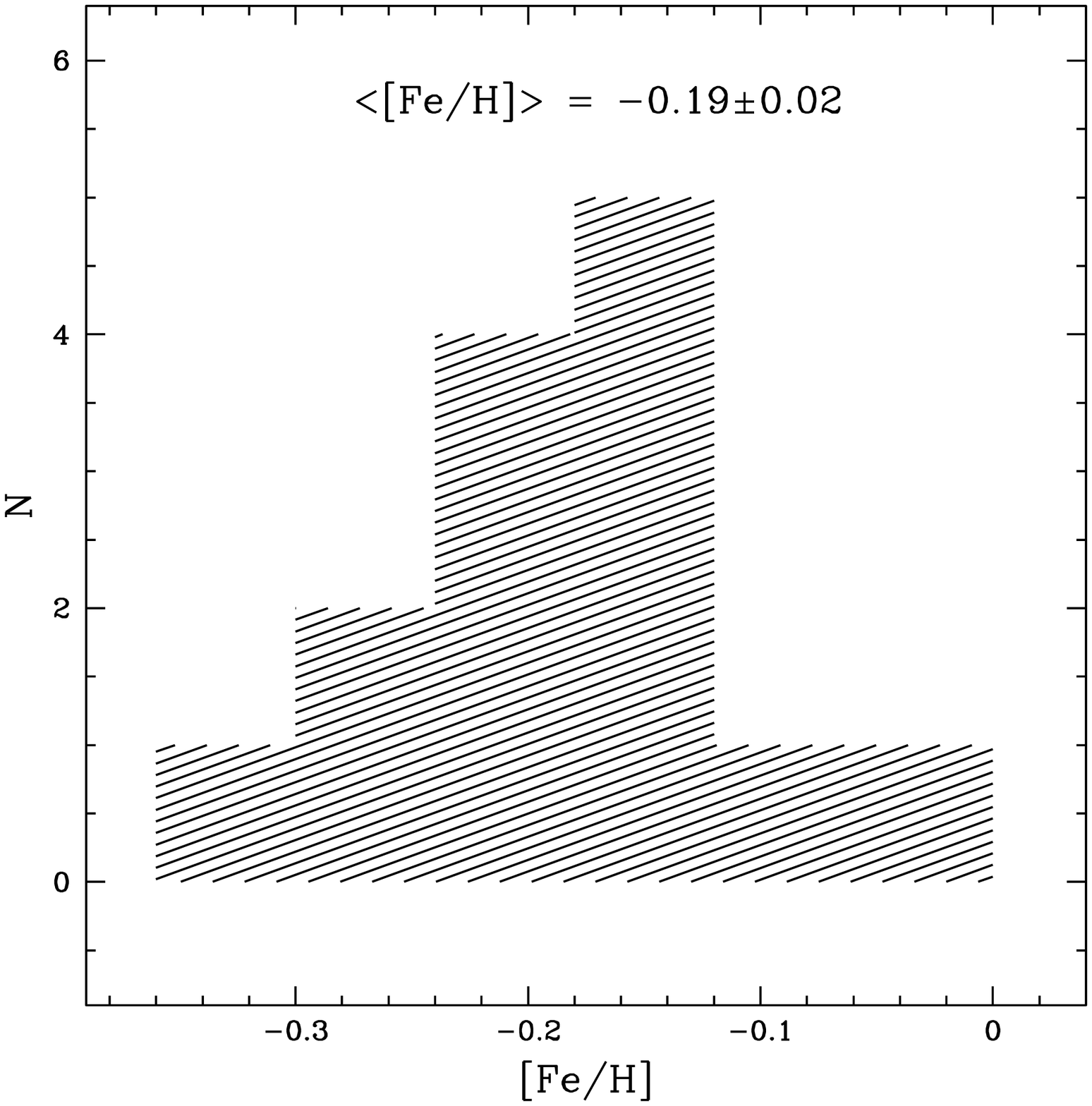}
\caption {Histogram of the metallicity distribution, as traced by [Fe/H],
for the observed giants in the Baade's window.
\label{histo}}
\end{figure}
We find no indication of striking iron or alpha abundance differences
between our M giant sample and the field and globular cluster bulge
K giants previously studied.  The observed composition requires no special 
population to be the progenitors of the M giants.  We cannot rule out the possibility that
alpha-enhanced stars are more likely to evolve into M giants, but our sample is
too small to comment on this possibility.

By observing both the local and bulge M giants with the
same instrumentation, we are able to show that the bulge
M giants are alpha enhanced,
when compared with stars in the Solar vicinity.  
Earlier work comparing Solar neighborhood and bulge M giants at low
resolution would not have been able to sort out the issues
connected with composition.  The low resolution spectroscopy of \citet{tfw90} and \citet{sharp90}
found that the bulge M giants have stronger TiO bands than their disk counterparts 
of the same temperature; we propose that alpha enhancement (raising Ti and O abundances)
is responsible for this.

Our survey of giant late type stars in Baade's window find their
metallicities to range in [Fe/H] between --0.4 and +0.0 dex,
(see Fig.~\ref{histo}).  
Arp \citep{arp65} I-194 (more likely a K giant), is in common with the \citet{MWR04} sample:
very similar (within $\approx$0.1 dex) Fe, O, Mg and Si abundances and
somewhat lower (by $\approx$0.3 dex) Ti and higher (by $\approx$0.3dex)
Ca abundances have been obtained in our study.    It would be useful to obtain
more stars with both optical and infrared abundance analysis, but the infrared
approach works best for temperatures cooler than 4000K.  Oxygen abundances in
the optical are based on  the [OI] 6300.3\AA\ line, while our oxygen abundances
are derived from numerous OH lines, and comparison with the disk stars 
confirms the oxygen enhancement.
As larger samples are developed, it will
be important to compare and reconcile the optical and infrared abundance determinations, especially
as one pushes toward the Galactic center where only infrared spectroscopy is possible.

The metallicity distribution of the bulge stars in our survey
peaks at slightly subsolar metallicity, in agreement with
the giant K star distribution \citep{RMW00,MWR04} and the
recent, extensive photometric survey by \citet{zoc03}.
We note the apparent absence of M giants with
iron abundances above solar, while their low metallicity spread
compared to the significantly larger one measured among K giants
by \citet{RMW00,MWR04} is not surprising and we expect to 
find higher metallicity M giants as the sample size increases.
Only relatively metal rich stars reach the M spectral type, with those
near the red giant branch tip reaching
temperatures as cool as $\approx$3500~K.  
However, we find no correlation between spectral
type and metallicity (the M9 giants are not extraordinarily metal rich).
We also find among the M giants no stars well above the Solar iron
abundance.  One of the goals of a larger survey would be to determine
whether there is a genuine deficiency of metal rich stars in the M giant
population.
Do such stars evolve through the M giant phase too rapidly
to be found in this small sample?  It is conceivable that higher metallicity
results in greater mass loss, and that the the most metal rich stars evolve
to the UV-bright AGB-manqu\'e phases.

\begin{figure}
\plotone{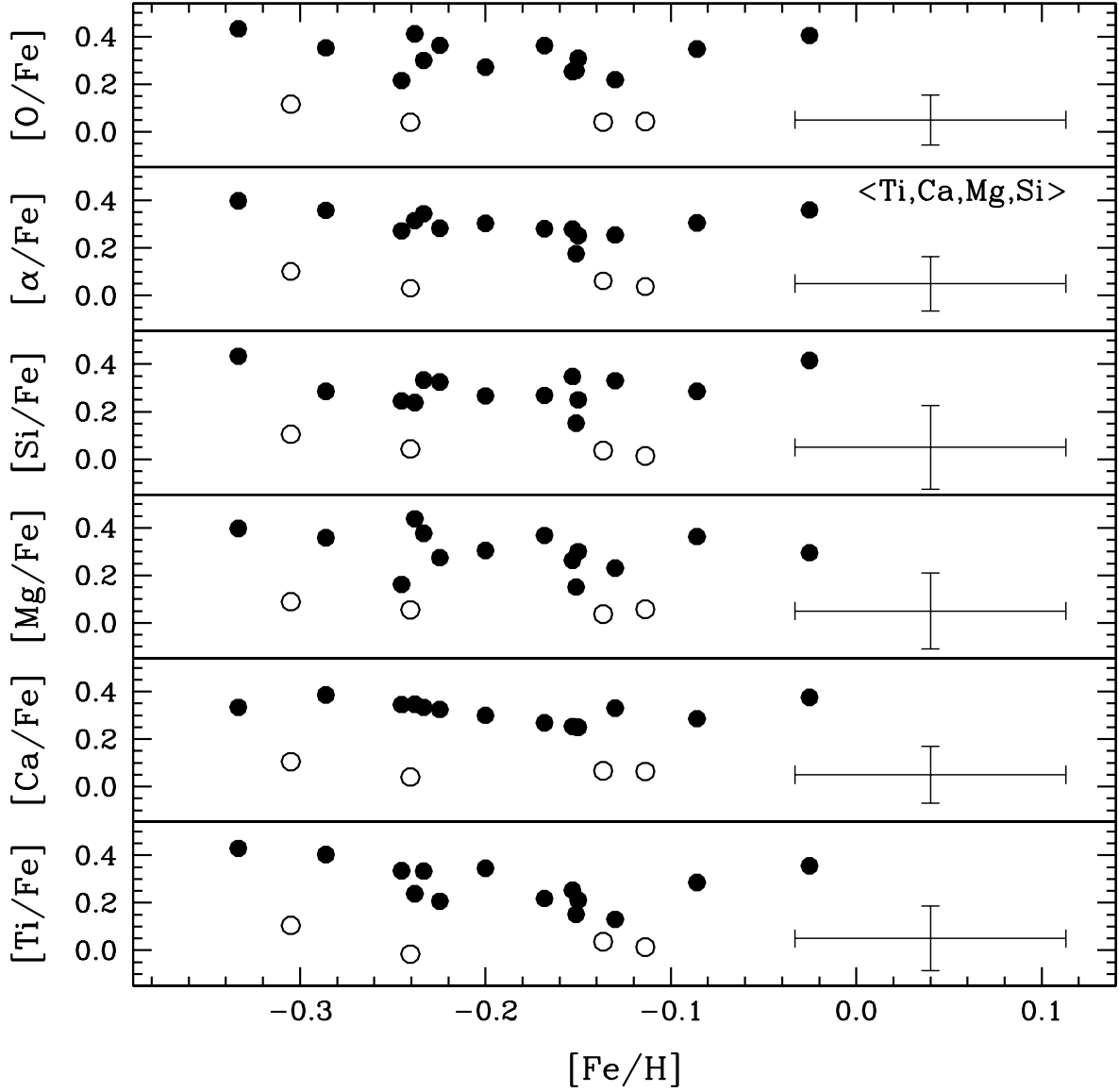}
\caption{$\alpha$-element over iron abundance ratios
as a function of [Fe/H] for the observed giants in the Baade's window
(full dots) and in the solar neighborhood (empty circles).
Typical error bars are plotted in the bottom-right corner of each panel.
\label{alpha}}
\end{figure}

\begin{figure}
\plotone{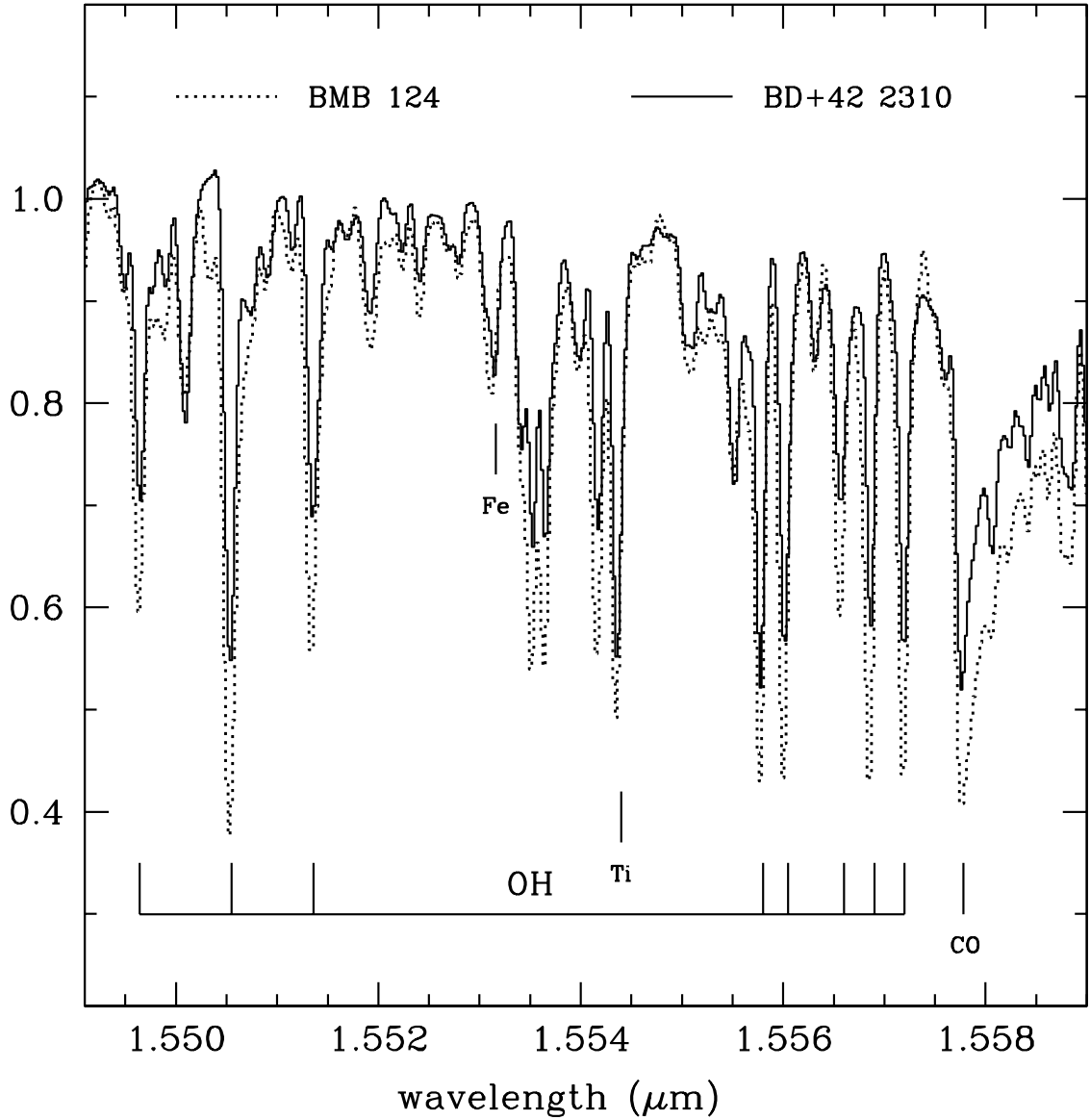}
\caption {A region of the NIRSPEC H--band spectrum around 1.555~$\mu$m of
one Baade's window (BMB124) and one solar neighborhood (BD+42~2310) M giant
with very similar photospheric parameters and metallicity.
BMB124 show much deeper OH and Ti lines compared to BD+42~2310, indicating
a bulge [$\alpha$/Fe] enhancement compared to the solar value.
\label{conf}}
\end{figure}

Our survey also shows a homogeneous $\alpha$-enhancement by $\approx$+0.3 dex
up to solar metallicity, without significant differences among the various
$\alpha$-elements (see Fig.~\ref{alpha}), while
the four solar neighborhood M giant stars in our control sample are consistent
with solar [$\alpha$/Fe] values.
As a further probe of the different degree of $\alpha$ enhancement between the
Galactic bulge and the disk, Fig.~\ref{conf} compares the spectra of a bulge
and a solar neighborhood M giant star with almost identical photospheric
parameters and iron abundances: the OH and Ti lines of the bulge giant
are clearly deeper.  

While we are dealing with a small sample of M giants, it is fair to say that
if they were evolved from a younger sub population of the bulge, we would
expect these stars to have Solar $[\alpha /Fe]$.

\begin{figure}
\plotone{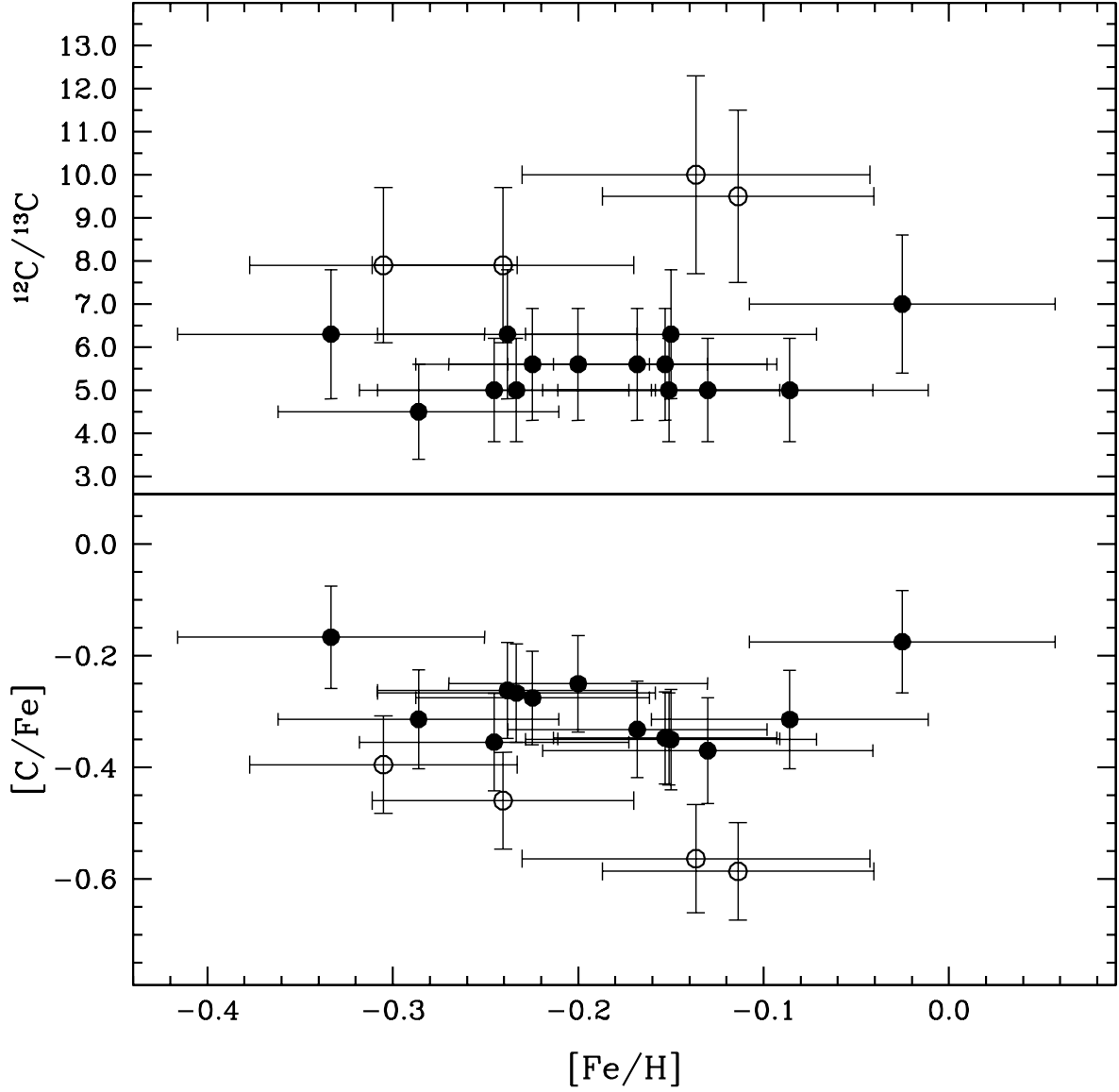}
\caption {[C/Fe] and $\rm ^{12}C/^{13}C$ abundance ratios as a function of
[Fe/H] for the observed giant stars in the Baade's window (full dots) and in
the solar neighborhoods (empty circles).
\label{carbon}}
\end{figure}
We have also derived carbon abundances for all of the observed stars from
analysis of the CO bandheads.
Fig.~\ref{carbon} shows the [C/Fe] and $\rm ^{12}C/^{13}C$ abundance ratios
as a function of the iron abundance, which
provide major clues (the latter in particular, see \citet{ori03}
and references therein) to the efficiency of the mixing processes
in the stellar interiors during the evolution along the RGB.
We find some degree of [C/Fe] depletion with respect to solar values
and very low $^{12}$C/$^{13}$C$\le$10, as
also measured in metal-poor halo giants both in the field and in
globular clusters \citep[see e.g.][ and reference therein]{ss91,sh96,gra00},
as well as in bulge clusters \citep{orc02,sh03,ori04}.

The classical theory \citep[][ and references therein]{ib67,ch94}
predicts a decrease of $^{12}$C/$^{13}$C$\approx40$ after
the first dredge--up, the exact amount mainly depending on the chemical
composition and the extent of the convective zone.
Additional mixing mechanisms due to further {\it cool bottom processing}
\citep[see e.g.][]{ch95,dw96,csb98,bs99,wdc00}
can explain much lower $\rm ^{12}C/^{13}C$ values, as those measured
in the upper RGB stars.

\acknowledgments
R. Michael Rich acknowledges support from 
grant AST-0098739 from the National Science Foundation.
Livia Origlia acknowledges financial support by the Agenzia Spa\-zia\-le Ita\-lia\-na (ASI) and
the Ministero dell'Istru\-zio\-ne, Universit\`a e Ricerca (MIUR).

\end{document}